# Flow Transitions and Mapping for Undulating Swimmers

Muhammad Saif Ullah Khalid[1, 2], Junshi Wang[3], Haibo Dong[3], Moubin Liu[1, 2*]

[1] Institute of Ocean Research, Peking University, Beijing, People's Republic of China
[2] Key State Laboratory of Turbulence and Complex Systems, Department of Mechanics and Engineering Science, Peking University, Beijing, People's Republic of China
[3] Department of Mechanical and Aerospace Engineering, University of Virginia, Charlottesville, 22904, VA, USA

* Corresponding Author, Email: mbliu@pku.edu.cn

Natural swimmers usually perform undulations to propel themselves and perform a range of maneuvers. These include various biological species ranging from micro-sized organisms to large-sized fishes that undulate at typical kinematic patterns. In this paper, we consider anguilliform and carangiform swimming modes to perform numerical simulations using an immersed-boundary methods based computational solver at various Reynolds number (Re) regimes. We carry out thorough studies using wavelength and Strouhal frequency as the governing parameters for the hydrodynamic performance of undulating swimmers. Our analysis shows that the anguilliform kinematics achieves better hydrodynamic efficiency for viscous flow regime, whereas for flows with higher Re, the wavelength of a swimmer's wavy motion dictates which kinematics will outperform the other. We find that the constructive interference between vortices produced at anterior parts of the bodies and co-rotating vortices present at the posterior parts plays an important role in reversing the direction of Benard-von Karman vortex street. Since most of the thrust producing conditions appear to cause wake deflection; a critical factor responsible for degrading the hydrodynamic efficiency of a swimmer, we discuss the underlying mechanics that would trigger this phenomenon. We demonstrate that the choice of kinematic and flow conditions may be restricted for the natural swimmers due to their morphological structures, but our findings provide a guideline on their selection for bio-inspired underwater vehicles.

## 1. Introduction

Since the idea designing bio-inspired underwater robots was conceived, numerous research efforts (Lauder & Tytell, 2006; Akhtar, et al., 2007; Liu, et al., 2017) are underway to explain the hydrodynamics of natural biological species that make them agile and efficient and help them perform complex maneuvers. Most of the aquatic species employ a wavy motion; known as undulation, along their bodies to control the flows in their vicinity and propel themselves. Their kinematic profiles are usually classified into anguilliform, carangiform, subcarangiform, and thunniform (Sfakiotakis, et al., 1999; Lauder, 2015). Due to their adoption by a wide range of fish species, the first two kinds of motions and their hydrodynamic effects are the subject of our present work. It is well known that anguilliform swimmers has more than one wavelengths along their bodies that means the wavelength is generally shorter than their body-lengths (Sfakiotakis, et al., 1999; Lauder & Tytell, 2006). Contrary to that, the wavelengths are greater than the body-lengths in case of carangiform swimmers (Sfakiotakis, et al., 1999; Videler & Wardle, 1991). It is still an open question that whether such wavy motions are performed due to their physiological constraints or the purpose is to achieve certain hydrodynamic performance. Which type of undulation gives more hydrodynamic advantage is also an important consideration from the perspective of designing efficient underwater bio-inspired robots. Up to the author's knowledge, there are very limited efforts carried out in this direction. For instance, Borazjani and Sotiropoulos (2010) compared the propulsion metrics of anguilliform and carangiform swimmers for mackerel and eel type bodies. Their numerical simulations revealed that the anguilliform kinematics performed better not only in terms of obtaining higher velocities but also greater swimming efficiency for lower and mid-range Reynolds number flows; Re = $10^2$ and $10^3$. However, their carangiform swimmer dominated over the other one in terms of attaining higher hydrodynamic parameters in the inertial regime;Re = ∞. They used wavelengths, nondimensionalized by the body-lengths, of the undulating motion as 0.642 and 0.95 for the anguilliform and carangiform swimmers, respectively. After that, Maertens et al. (2017) compared the quasi-steady efficiencies of anguilliform and carangiform swimming modes with a unit wavelength for a self-propelling NACA-0012 shaped foil and found that anguilliform type undulation showed up as the more efficient one, but both these swimming modes achieved their optimum performance at the same excitation frequency. Besides, a major aim of their study was to explore an optimum gait in order to enhance the propulsive efficiency. Earlier in a similar context, Kern and Koumoutsakos (2006) conducted numerical simulations through an evolutionary algorithm to obtain higher efficiencies for an anguilliform swimmer. Their investigations uncovered that the swimmers modified their gait patterns during swimming in order to attain their hydrodynamic objectives. Next, van Rees et al. (2013) attempted to optimize the geometric shape of an anguilliform swimmer and observed that their engineered geometrical shape was able to outperform the naturally shaped larval zebrafish. These conclusions lead us to raise questions and understand important relevant features about the suitability of wavy kinematics for natural swimmers to fulfil their hydrodynamic requirements. This can be understood by performing thorough quantitative and qualitative analyses for a wider range of kinematic and flow parameters. This strategy will help us comprehend flow transitions and suitability of different kinematic parameters under various flow conditions for natural undulating swimmers.

Due to the level of complexities involved with this, it is also a common practice to use harmonically oscillating foils; as a simplified model for bio-inspired propulsive mechanisms, to quantify their hydrodynamic performance and to investigate the underlying governing flow mechanics (Anderson, et al., 1998; Godoy-Diana, et al., 2008; Godoy-Diana, et al., 2009; Deng, et al., 2016; Floryan, et al., 2017; Cleaver, et al., 2012; Lagopoulos, et al., 2019). This traditional strategy helped established various similarities with the physics associated with the actual natural swimming. Godoy Diana et al. (2009; 2008), Deng et al. (2016), and Lagopoulos et al. (2019) developed transition maps for flows past simply oscillating foils in the phase planes constituted by oscillation amplitudes and Strouhal number; a nondimensional measure of excitation frequency. A common observation was the transformation of the Benard von Karman vortex street (BvKVS) into reverse Benard von Karman vortex street (RBvKVS) preceding the production of thrust for the body. This corresponds to the insufficient support to a body provided by RBvKVS to overcome its frictional drag at lower values of the kinematic parameters. Another important aspect is the emergence of jet deflecting on either side of the body, causing a net lateral force. It is still unknown whether such transitions can be observed for wavy motion of a body because this motion does not usually allow the flow to separate from the swimmer's body during its travelling to the posterior parts. Using numerical simulations through a high-fidelity immersed-boundary methods based computational solver, our present study aims at carrying out a systematic investigations on various important aspects related to undulating swimming which include; (1) do we observe such flow transitions for undulating swimmers?, (2) how are the transitions influenced by viscous and inertial effects introduced by variations in Reynolds number?, (3) do the transitions depend on the kinematic waveform of the undulating swimmer?, (4) how do anguilliform and carangiform swimmers hydrodynamically perform in comparison to each other given their kinematic profiles is influenced by the wavelength and excitation frequency?, (5) what are the differences between their flow physics?, and (6) do these undulating swimmers also experience wake deflection? The change in Reynolds number (Re) addresses two kinds of variations. Considering the body-length as the characteristic dimension, the variation in hydrodynamic performance metrics help us understand how fishes may adopt different kinematic parameters including the wavelength and frequency at different stages of its life because the body-length is a measure of age and growth. On the other hand, given the characteristic velocity as the controlling parameter for variations in Re, it also provides an answer on the required variations in their gaits if they want to swim faster or more efficiently.

## 2. Numerical Methodology

*Geometry and Kinematics of Swimmers*

In our present work, we employ NACA0012 foil to model the swimmers' bodies where its chord represents the spine of a swimmer at the time of their static equilibrium. Two types of wavy kinematic modes are considered here; anguilliform and carangiform. Eels are classified as the representatives for anguilliform swimmers which undulates a large portion of their bodies to move, whereas those belonging to carangiform class include fish species having distinct caudal fins attached with their bodies (Lauder & Tytell, 2006). With the chord representing the backbone of a fish, the carangiform amplitude profile is given by the following relation (Khalid, et al., 2016; Khalid, et al., 2018);

$$A\left(\frac{x}{L}\right) = 0.02 - 0.0825\left(\frac{x}{L}\right) + 0.1625\left(\frac{x}{L}\right)^2 ; 0 < \frac{x}{L} < 1$$

where $A\left(\frac{x}{L}\right)$ denotes the local amplitude at a given spatial position along the fish body; nondimensionalized by its total length ($L$). The coefficients $a_o$, $a_1$, and $a_2$ are calculated based on the data provided for a steadily swimming saithe fish (Videler, 1993) where the local amplitudes are $A(0) = 0.02$, $A(0.2) = 0.01$, and $A(1.0) = 0.10$.

For the anguilliform swimming mode, we model the amplitude envelope using the following equation (Tytell & Lauder, 2004; Borazjani & Sotiropoulos, 2010; Maertens, et al., 2017).

$$A\left(\frac{x}{L}\right) = 0.0367 + 0.0323\left(\frac{x}{L}\right) + 0.0310\left(\frac{x}{L}\right)^2 ; 0 < \frac{x}{L} < 1$$

The undulatory kinematics takes the following form in both the cases.

$$y_o\left(\frac{x}{L}, t\right) = A\left(\frac{x}{L}\right) \cos\left[2\pi\left(\frac{x}{\lambda} - ft\right)\right]$$

Here, the factor $2\pi/\lambda$ defines the wave-number ($k$) for the waveform of the kinematic profile along the swimmer's body and we introduce the non-dimensional wavelength as $\lambda^* = \lambda/L$ to conduct our parametric study. Also, we limit the maximum amplitude of the trailing-edge to $0.10$ for the two swimmers. We display the amplitude envelope for both kinds of swimmers in Figure 1(a), whereas their respective wavy kinematics during an oscillation cycle are presented in Figure 1(b) and (c).

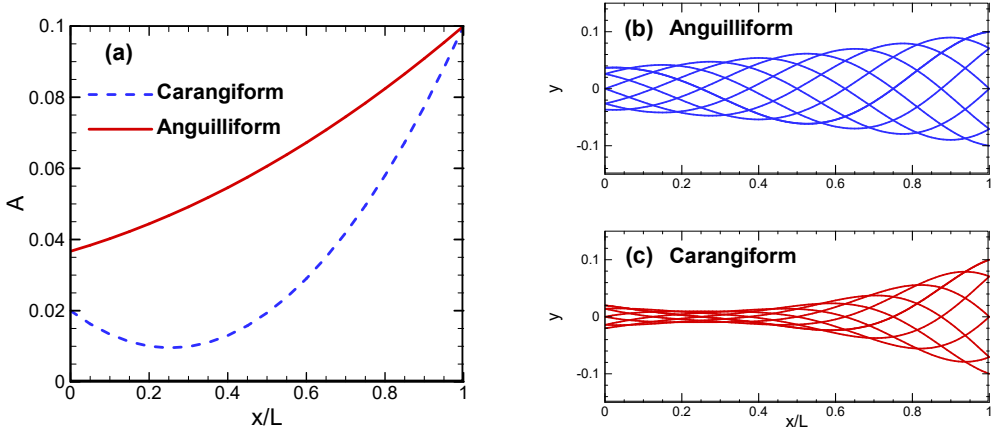

**Figure 1:** (a) Amplitude envelope for carangiform and anguilliform undulations, (b) backbone motion of an anguilliform swimmer, and (c) backbone motion of a carangiform swimmer

The swimmer is placed in rectangular virtual tunnel as shown in Figure 2 where the dimensions of the flow domain are 40L × 15L. The purpose is not to draw the figure to a scale but to show the geometric details only.

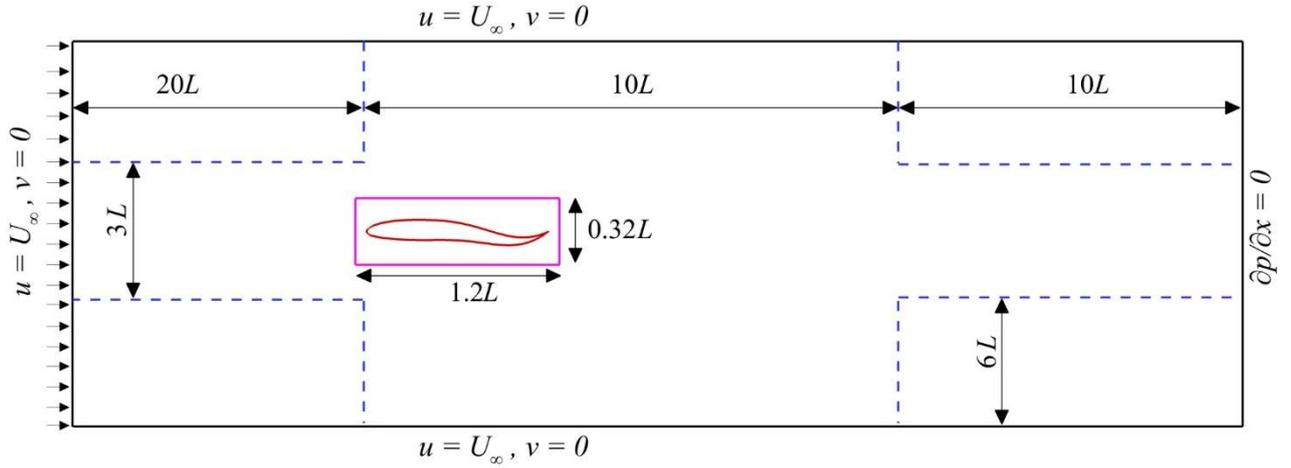

**Figure 2 :** Flow domain and the boundary conditions

*Computational Solver*

We perform two dimensional (2D) numerical simulations for the swimmers while neglecting its three-dimensional flow effects; an effective computational strategy to characterize salient features of the swimming dynamics and flow physics (Khalid, et al., 2016; Gazzola, et al., 2014). Following non-dimensional forms of the continuity and incompressible Navier-Stokes equations constitute the mathematical model for the fluid flow.

Continuity Equation:
$$\frac{\partial u_i}{\partial x_i} = 0$$

Navier-Stokes Equations:
$$\frac{\partial u_i}{\partial t} + \frac{\partial u_i u_j}{\partial x_j} = -\frac{\partial p}{\partial x_i} + \frac{1}{Re}\frac{\partial^2 u_i}{\partial x_i \partial x_j}$$

where the indices $\{i, j\} = \{1,2,3\}$, $x$ shows a Cartesian direction, $u$ denotes the Cartesian components of the fluid velocity, $p$ is the pressure, and Re represents the Reynolds number; defined as $Re = U_\infty L/\nu$. Here, we keep $U_\infty = 1$ m/sec, $L = 1$ m, and vary $\nu$ to change Re.

We solve the described governing model for fluid flow using a Cartesian grid-based sharp-interface immersed boundary methods (Mittal, et al., 2008) where the spatial terms are discretized using a second-order central difference scheme and a fractional-step method is employed for time marching. This makes our solutions second-order accurate in both time and space. We utilize the Adams-Bashforth and implicit Crank-Nicolson schemes for respective numerical approximation of convective and diffusive terms. The prescribed wavy kinematics is enforced as a boundary condition for the swimmers. We impose such conditions on immersed bodies through a ghost-cell procedure (Mittal, et al., 2008) that is suitable for both rigid and membranous body-structures. Further details of this solver and its employment to solve numerous bio-inspired fluid flow problems is available in Ref. (Fish, et al., 2016; Liu, et al., 2017; Wang, et al., 2019).

As displayed in Figure 2, we employ Dirichlet boundary conditions for the left, top, and bottom boundaries for flow velocities, whereas Neuman conditions are used at the outflow boundary on the right side. The dashed lines show the region with high mesh density in order to adequately resolve the flow features in the wake. The rectangular box encompassing the swimmer's body shows the region where we use even higher mesh density to capture the boundary layer flows at higher Reynolds numbers. Before setting up our actual simulations for this research work, we perform thorough tests for the convergence of grid size and time-step size ($\Delta t$) for the undulation frequency ($f$) 0.5 Hz and Re = 5000 for a carangiform swimmer modeled with a NACA0012 foil. In Table 1, we show the details for the grid-convergence study. Here $n_x \times n_y$ denotes the grid size inside the innermost highly dense mesh region. We define $\Delta t = \tau/N$, where $N$ is number of time-steps permitted for each oscillation cycle time-period ($\tau$). Based on our findings, we choose to continue for our simulations with the settings of test 7, i.e. $N_x \times N_y = 1537 \times 897$ with $N = 2000$.

**Table 1:** Details of our grid size and time-step size independence study

| Serial No. | Mesh size ($n_x \times n_y$) | No. of time-steps in each oscillation cycle ($N$) | Average Thrust Coefficient $\overline{C_D}$ |
|---|---|---|---|
| 1 | 64 × 32 | 2000 | 0.1187 |
| 2 | 128 × 64 | 2000 | 0.1134 |
| 3 | 192 × 96 | 2000 | 0.0972 |
| 4 | 256 × 128 | 2000 | 0.0992 |
| 5 | 320 × 160 | 500 | 0.0996 |
| 6 | 320 × 160 | 1000 | 0.0992 |
| 7 | 320 × 160 | 2000 | 0.0983 |
| 8 | 320 × 160 | 4000 | 0.0970 |
| 9 | 384 × 192 | 2000 | 0.0982 |

*Performance Parameters*

To check the hydrodynamic performance of a swimmer, we use the propulsive efficiency defined by Deng et al. (2007) and Dong and Lu (2007) where it is defined as the ratio between the power used by the swimmer to overcome the drag force and that consumed by the swimmer to keep its body oscillating to follow its kinematic profile.

$$\eta = -\overline{P_D}/\overline{P_S}$$
$$P_D = F_D U_\infty$$

where $F_D = 0.50 C_D \rho L U_\infty^2$ is the drag being experienced by the swimmer due to the on-coming fluid. It is important to notice that $C_T = -C_D$. As each point on a swimmer's body oscillates in the normal direction to the free-stream fluid flow, the power expended by a swimmer to perform the wavy motion is computed as;

$$P_S = \oint (\overline{\overline{\sigma}} \cdot \boldsymbol{n}) \cdot \boldsymbol{V} ds$$

where $\oint$ is the surface integral operator, $\overline{\overline{\sigma}}$ denotes the stress tensor, $\boldsymbol{n}$ shows the normal vector to the body surface, and $\boldsymbol{V}$ shows the fluid velocity vector adjacent to the swimmer's body. An important parameter in analyzing hydrodynamics of bio-inspired locomotion is the non-dimensional Strouhal frequency ($f^*$) that is defined as;

$$f^* = \frac{2A_{max}f}{U_o}$$

Here, $A_{max}$ is the maximum amplitude traversed by a swimmer's tail that is treated as a measure of the vortical wake-width. In this paper, we compute all the time-averaged quantities; denoted by a "¯" over a symbol, using the last 5 oscillations cycles, whereas the solutions gets to their steady-state within 2-3 initial cycles.

*Governing Parametric Space*

We explain the governing variables in this study and their relevant specifications in Table 2. In order to investigate the fluid dynamics and carry out a detailed quantitative analysis, we use three different values of Reynolds number; $10^2$, $10^3$, and $5 \times 10^3$, that represent the viscous, transitional, and inertial regimes, respectively, in bio-inspired swimming related phenomena (Borazjani & Sotiropoulos, 2010). The range of $\lambda^*$ covers the observations made by various biologists.

**Table 2:** Explanation of the governing variables and their specifications

| Parameters | Specifications |
|---|---|
| Geometry | NACA-0012 |
| Undulatory kinematics | Carangiform and Anguilliform |
| Re | $10^2$ (viscous), $10^3$ (transitional), and $5 \times 10^3$ (inertial) |
| $f^*$ | 0.10 - 0.80 |
| $\lambda^*$ | 0.50 - 1.5 |

## 3. Results & Discussion

Here, we present our qualitative analysis for hydrodynamic performance parameters of both the anguilliform and carangiform swimmers along with the vortex dynamics and the wake mechanics. This discussion also includes an illustration of the deflected wake phenomenon.

We begin our discussion with phase maps in $\lambda^* - f^*$ plane as shown in Figure 3 which shows three flow regimes and their transition points in the whole governing parametric space for both the anguilliform and carangiform swimmers. These plots clearly show the similarities and dissimilarities in their global flow dynamics. Now, it is interesting to observe that reverse Benard von Karman vortex street (RBvKVS) starts forming; the phenomenon generally used as an indicator for thrust production by a swimming body, before the swimmers start producing thrust. This was also observed and reported previously by Streitlien & Triantafyllou (1998), Godoy-Diana et al. (2008), Bohl & Koochesfahani (2009), and Lagopoulos et al. (2019) for airfoils undergoing simple harmonic oscillations; known as pitching or plunging. This happens because of the contribution from the momentum-surfeit wake in the downstream direction of the body is not enough to overcome the profile or frictional drag. For the viscous regime (Re = $10^2$), the anguilliform produces thrust for a broader parametric range and, even, it experiences jet deflection in its wake for the higher values present in the right-most corner of Figure 3a. Phase maps for the transitional flow regime (Re = $10^3$) in Figure 3c & d and those for the inertial regime (Re = $5 \times 10^3$) in Figure 3e & f display that the global dynamics for the two kinds of swimmers is quite similar. We present the quantification of their respective hydrodynamic parameters; $\overline{C_T}$, $\overline{P_S}$, and $\eta$, in Figure 4, Figure 5, and Figure 6 for viscous, transitional, and inertial flow regimes. For Re $= 10^2$, magnitudes of these performance indicators show an increasing trend for increasing values of both $\lambda^*$ and $f^*$. We keep the axes on the same scale for respective hydrodynamic parameters of both the bodies for the sake of comparison. Our anguilliform swimmer produces more amount of thrust as compared to its counterpart, but that comes at the cost of consuming more swimming power. It is apparent that $\overline{C_T}$ for the carangiform swimmer is not largely influenced by an increasing $\lambda^*$ at a specific value of $f^*$ as compared to the anguilliform swimmer where the impact of $\lambda^*$ on the axial force is more pronounced. The higher the wavelength, the greater the thrust it produces and the effect is more definite in the thrust producing regime as compared to that exhibiting drag production.

A comparison between the metrics of the two swimmers tells us that the anguilliform one outperforms its counterpart within the thrust production regime for Re $= 10^2$. It not only comes out to be a faster swimmer here (in terms of a greater thrust generation) but also appear as a more efficient swimmer as compared to the body adopting carangiform mode. An incremental change in $\lambda^*$ by 0.25 at $f^* = 0.70$ and 0.80 brings almost 130% and 66% higher thrust force, respectively, along with respective 90% and 36% enhancements in its efficiency for the anguilliform swimmer when it is seen in comparison with the carangiform swimmer. However, when it comes to the transitional and inertial flow regimes (Re $= 10^3$ and $5 \times 10^3$), we observe that the carangiform swimmer performs better not only in terms of higher thrust but also hydrodynamic efficiency for $\lambda^* \leq 0.75$ of the whole $f^*$ range under consideration here. While undulating with $\lambda^* = 1.0$, the anguilliform kinematics gives the swimmer a slight edge over the carangiform mode at $Re = 10^3$ where its $\overline{C_T}$ is, at most, 7% higher with over 10% increase in $\eta$ for all the frequencies. Nevertheless for $\lambda^* > 1.0$, we record more than 26% betterment in $\overline{C_T}$ and 25% improvement in $\eta$ for our anguilliform swimmer for all $f^*$ values. In the inertial regime, $\overline{C_T}$ of the body with anguilliform mode, for $\lambda^* > 1.0$ gets improved by more than 22% and its $\eta$ is enhanced by more than 15%. For $\lambda^* = 1.0$, $\overline{C_T}$ from the anguilliform type kinematics is less than that from the carangiform motion for $f^* < 0.60$, but it gives higher $\eta$ by more than 9% for all $f^*$. This quantitative analysis and its comparison with the previously reported observations (Sfakiotakis, et al., 1999; Lauder, 2015) indicates that the natural swimmers may not choose their optimal gaits due to their physiological constraints when they swim steadily. It also shows us an opportunity to introduce design considerations based on these above mentioned guidelines for unmanned underwater bio-inspired vehicles that may outperform the natural aquatic swimmers.

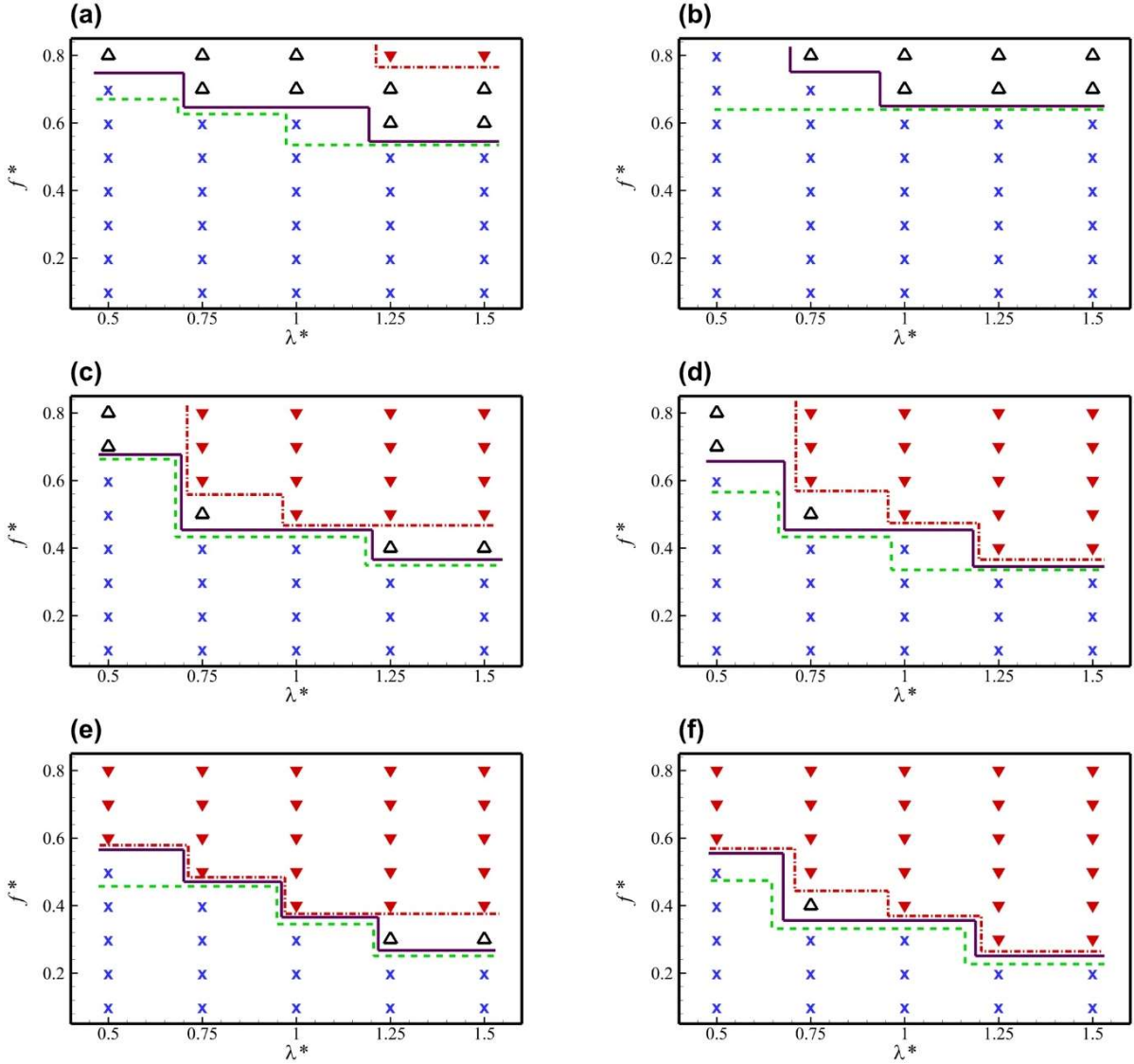

**Figure 3:** These Phase maps show flow regimes and their transitions for both the undulating swimmers. Here, the left (a, c, and e) and right columns (b, d, and f) correspond to the anguilliform and carangiform swimmers, respectively. The top row (a and b), middle (c and d), and bottom (e and f) rows present the characterization of flow regimes for $Re = 10^2$ (viscous regime), $Re = 10^3$ (transitional regime), and $Re = 5 \times 10^3$ (inertial regime). The symbols × (blue), $\Delta$ (black), and $\nabla$ (filled with red) represent the markers for parameters producing drag, thrust, and thrust with jet deflection, respectively. Moreover, Dashed (green) line is to show points for the transformation of BvKVS to its RBvKVS. Solid (violet) line demarks the thrust-generating parameters from those producing drag and the dotted (red) line separates the region for thrust production along with the jet deflection.

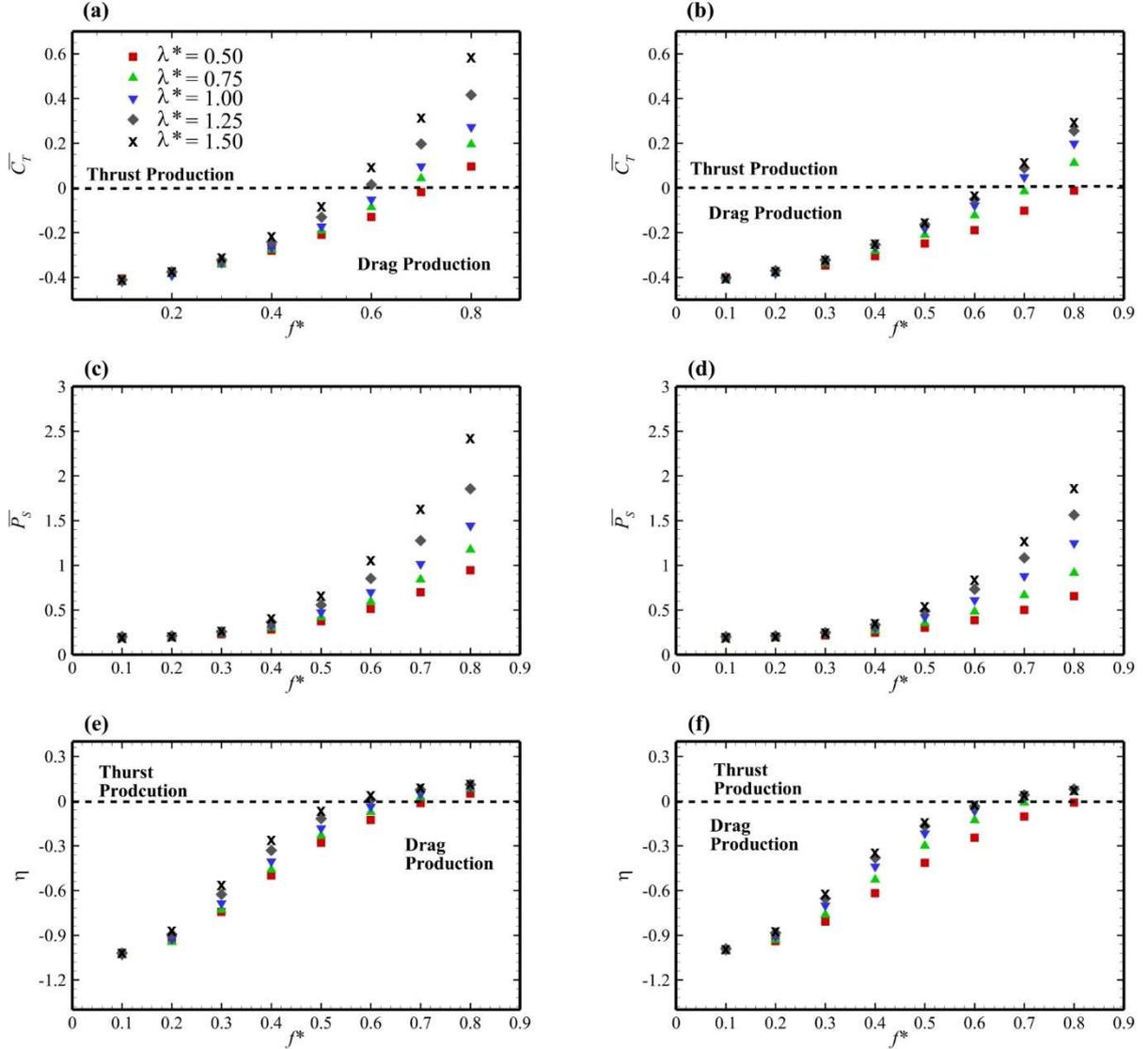

**Figure 4:** Time-averaged hydrodynamic parameters at Re $= 10^2$ for undulating swimmers where the left (a, c, and e) and right (b, d, and e) columns show the data for anguilliform and carangiform swimmers, respectively

To further comment on the individual performance, the anguilliform swimmer experiences a raise of 97% in $\overline{C_T}$, 25% in $\overline{P_S}$, and 58% in $\eta$ at $f^* = 0.70$ for each increment of $\Delta\lambda^* = 0.25$ on average. For For $f^* = 0.80$, these quantities are 59%, 27%, and 26%, respectively. Moreover, we find $\overline{C_T}$ and $\eta$ showing increases by large proportions by expanding on 25% more $\overline{P_S}$ at $f^* = 0.60$. Similarly, we observe an increase of 59% in $\overline{C_T}$, 21% in $\overline{P_S}$, and 31% in $\eta$ for $f^* = 0.70$ of the carangiform swimmer, whereas these respective quantities are 23% each for $\overline{C_T}$ and $\overline{P_S}$ for $f^* = 0.80$. However, the swimming efficiency of this swimmer decrease from 0.082 to 0.079 when $\lambda^*$ goes from 1.25 to 1.50. This data reveals the sensitivity of the hydrodynamic performance of both the swimmers for kinematic parameters while swimming in a viscous flow regime. Nevertheless, this also tells us that, by and large, the anguilliform kinematics performs better than the carangiform motion profile and helps the swimmer achieve better thrust and efficiency by expending a little more amount of power to perform its undulation. A plausible reasoning behind this observation may be related to the greater amount of body's oscillations this swimmer undergoes, thus leading to the production of more frictional drag as well.

Coming to the time-averaged hydrodynamic performance parameters in the transitional regime, i.e. for Re = $10^3$, $\overline{C_T}$ of both the swimmers enhances tremendously. This change in Reynolds number

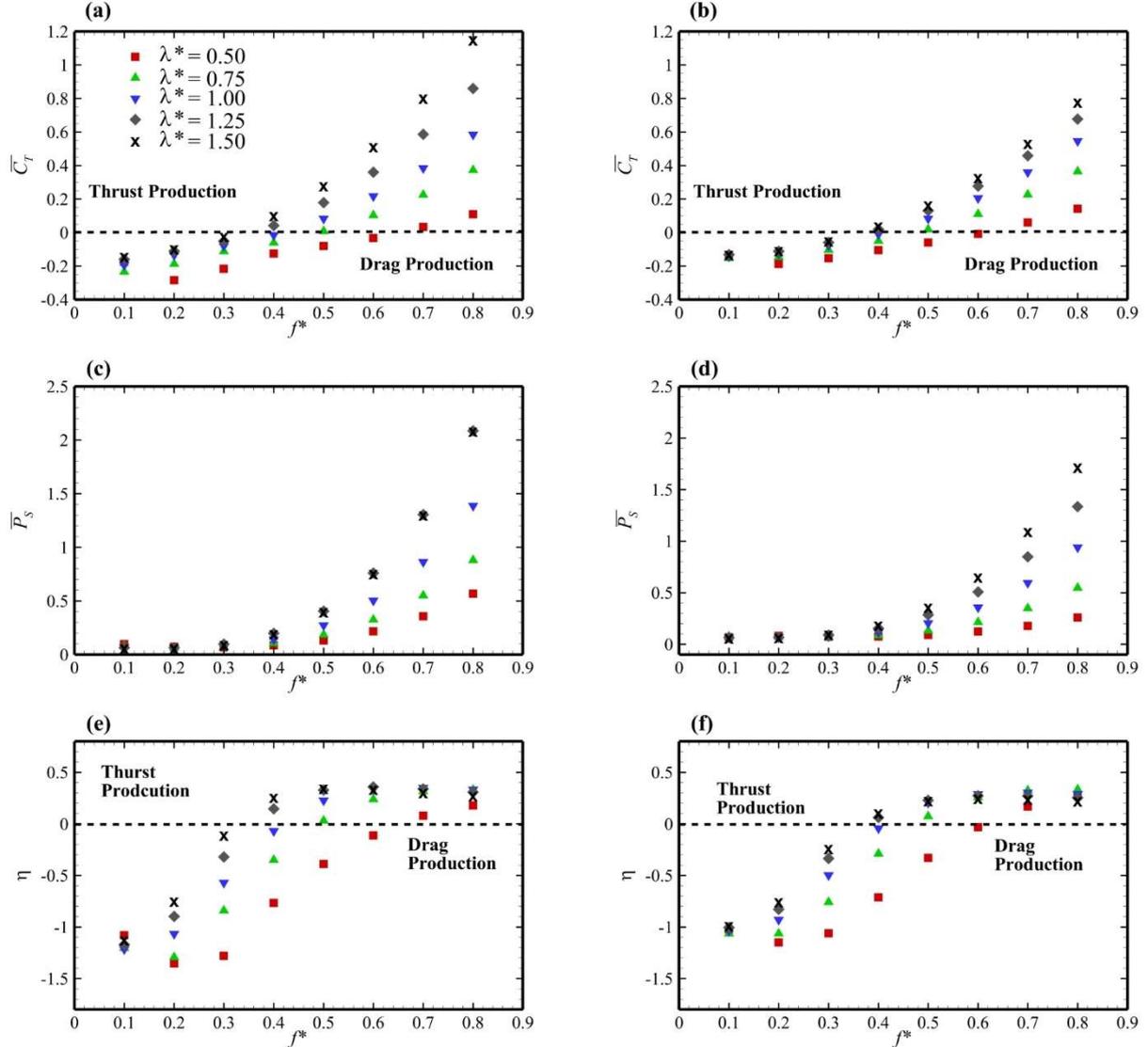

**Figure 5:** Time-averaged hydrodynamic parameters at Re = $10^3$ for undulating swimmers where the left (a, c, and e) and right (b, d, and e) columns show the data for anguilliform and carangiform swimmers, respectively

brings significant increments in $\overline{C_T}$ and $\eta$. We observe that $\overline{C_T}$ comes to the positive range for most of the kinematic parametric space. Considering the parameters where the anguilliform swimmer produces thrust by employing $\lambda^* = 0.50$, an average increments of 15% is seen with the reduction of 67% in $\overline{P_S}$ with an enhancement of $\eta$ manifolds. A similar trend is shown by the carangiform swimmer as well where it brings itself in the thrust generation domain by saving 60% power. It is also obvious that increase in $\lambda^*$ at Re = $10^3$ affects the thrust production positively for both the swimmers, but this costs more by requiring them to consume more power. An important feature here is the achievement of an optimum hydrodynamic efficiency at $f^* = 0.50$ for both swimmers. When the anguilliform swimmer increases its $\lambda^* = 1.0$ to 1.50, its $\overline{C_T}$, $\overline{P_S}$, and $\eta$ increases 3.37, 2.20, and 1.53 times, respectively. These respective parameters for the body performing carangiform kinematics are 2.00, 1.78, and 1.13 times. The purpose of reporting these variations is to explore a reasoning for the choice of wavelengths by both kinds of swimmers. Our present quantitative analysis reveals that the choice of kinematic parameters to obtain good hydrodynamic performance is dependent on the flow conditions provided a swimmer's physiology and morphological structure allows it to propel itself with those parameters. Our findings become more important since there is no existing comparative data in literature to the authors' knowledge that provides a detailed insight for understanding the kinematic mechanisms which may be opted to gain higher swimming performance.

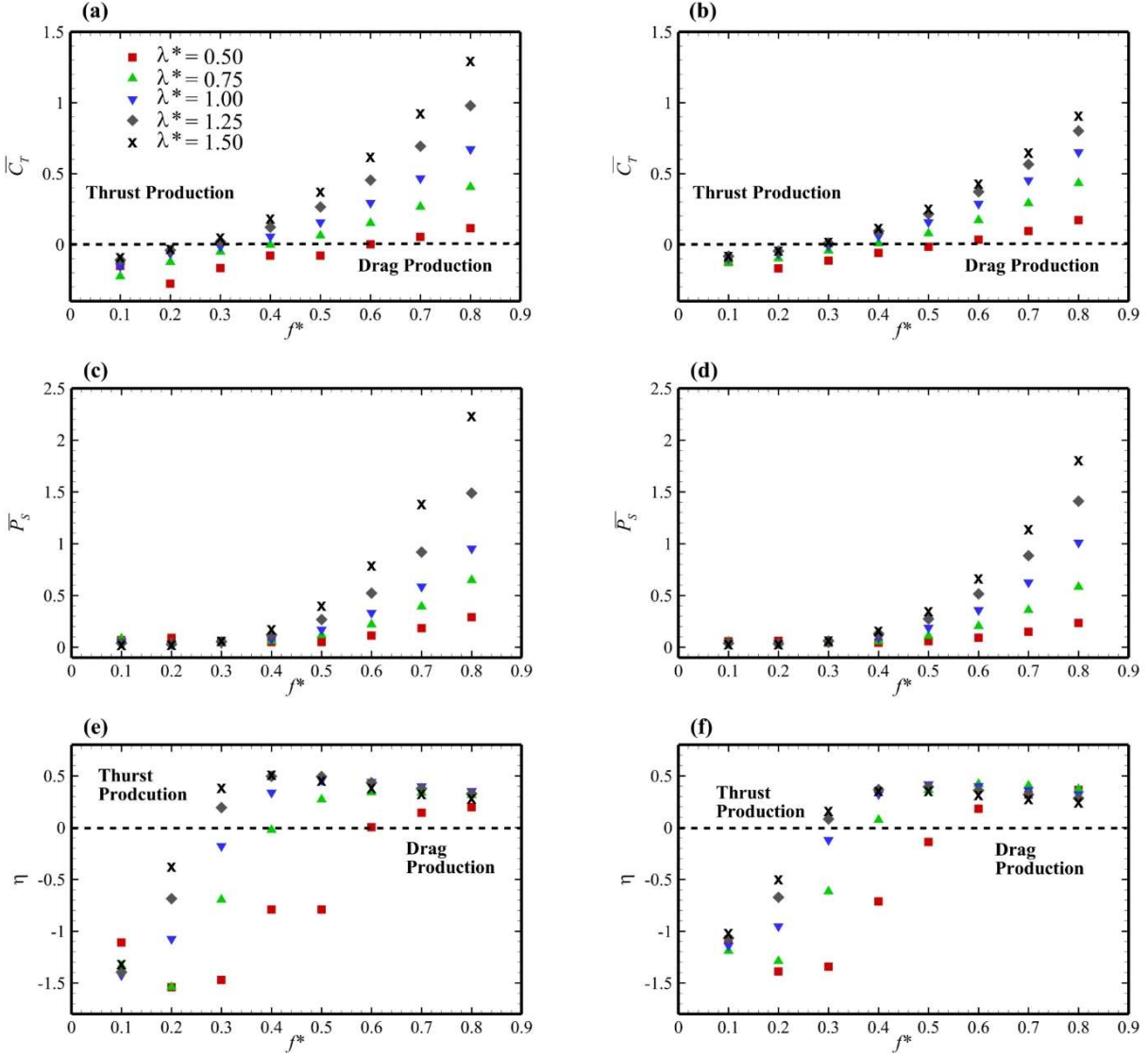

**Figure 6:** Time-averaged hydrodynamic parameters at $Re = 5 \times 10^3$ for undulating swimmers where the left (a, c, and e) and right (b, d, and e) columns show the data for anguilliform and carangiform swimmers, respectively

Next, we notice that trends in the variations of $\overline{C_T}$, $\overline{P_S}$, and $\eta$ for the inertial regime ($Re = 5 \times 10^3$), shown in Figure 6 are quite similar to those witnessed at $Re = 10^3$. It is important to highlight that the optimum efficiency for the most values of $\lambda^*$ shows up at or near $f^* = 0.40$ for both swimmers. However, the maximum $\eta$ appears at $f^* = 0.50$ when both swimmers undulate with $\lambda^* = 0.75$ and $1.00$. The hydrodynamic efficiency does not increase much with an increasing $\lambda^*$ at $f^* = 0.40$ and $0.50$; however, $\overline{C_T}$ increases significantly at the cost of higher amount of work done by the bodies to carry out its wavy kinematics. In case of carangiform swimmer, greater $\lambda^*$ decreases $\eta$ for $f^* \geq 0.60$. A similar pattern is observed for its counterpart as well. Furthermore, we notice that $\overline{C_T}$ and $\overline{P_S}$ increases when a body swims with a higher $\lambda^*$. Nevertheless, how does this choice impact its efficiency depends on the Reynolds number as well. The emergence of optimum Strouhal frequency to maximize $\eta$ is also dependent on the wavelength of the kinematic wave traversing through the body for all the flow regimes. Regardless of their relative magnitudes, we plot the optimum efficiency points in $f^* - \lambda^*$ plane for $Re = 10^3$ and $5 \times 10^3$ in Figure 7. Here, we consider only those points where $\eta$ starts degrading for higher $f^*$ with specific values of $\lambda^*$ for both the swimmers. Quite interestingly, both the swimmers follow, in general, almost similar linear patterns in this phase plane. These findings for higher range of $\lambda^*$ corroborate with those reported earlier (Anderson, et al., 1998; Triantafyllou, et al., 1993; Floryan, et al., 2017) for foils performing simple oscillations; namely heaving and pitching.

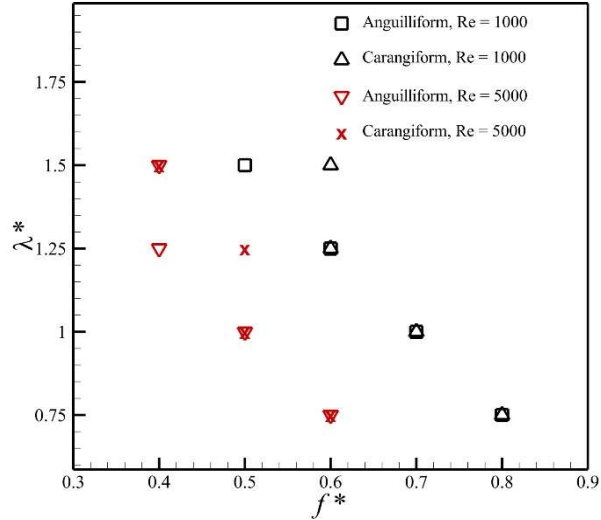

**Figure 7:** Optimum efficiency parameters in $f^* - \lambda^*$ phase plane

The corresponding values of the maximum $\eta$ for the parameters shown in Figure 9 are given in Table 3 that helps us discover some interesting characteristics about undulating swimmers. Except for $\lambda^* = 0.75$, the anguilliform shows higher optimum values of $\eta$ for both the transitional and inertial flow regimes.

**Table 3:** Maximum $\eta$ for the optimum kinematic parameters for our anguilliform and carangiform swimmers

| $Re$ | $\lambda^*$ | Anguilliform | | Carangiform | |
|---|---|---|---|---|---|
| | | $f^*$ | $\eta$ | $f^*$ | $\eta$ |
| $10^3$ | 0.75 | 0.80 | 0.3292 | 0.80 | 0.3341 |
| | 1.00 | 0.70 | 0.3492 | 0.70 | 0.3028 |
| | 1.25 | 0.60 | 0.3576 | 0.60 | 0.2732 |
| | 1.50 | 0.60 | 0.3391 | 0.60 | 0.2521 |
| $5 \times 10^3$ | 0.75 | 0.60 | 0.3422 | 0.60 | 0.4213 |
| | 1.00 | 0.50 | 0.4587 | 0.50 | 0.4175 |
| | 1.25 | 0.40 | 0.4963 | 0.50 | 0.3949 |
| | 1.50 | 0.40 | 0.5196 | 0.40 | 0.3624 |

**Table 4:** Kinematic parameters at $Re = 10^2$ for which the anguilliform swimmer produces thrust but the carangiform one does not do this.

| Case | $\lambda^*$ | $f^*$ | Anguilliform | | | Carangiform | | |
|---|---|---|---|---|---|---|---|---|
| | | | $C_{DF}$ | $C_{DP}$ | $C_T$ | $C_{DF}$ | $C_{DP}$ | $C_T$ |
| 1 | 0.50 | 0.80 | 0.6862 | -0.7817 | 0.0955 | 0.6129 | -0.6009 | -0.0119 |
| 2 | 0.75 | 0.70 | 0.7073 | -0.75 | 0.0427 | 0.6317 | -0.6162 | -0.0154 |
| 3 | 1.25 | 0.60 | 0.7051 | -0.7194 | 0.0143 | 0.6518 | -0.6017 | -0.0502 |
| 4 | 1.50 | 0.60 | 0.7514 | -0.8482 | 0.0968 | 0.6863 | -0.656 | -0.0303 |

Despite the similarities in their performance, there exist some important differences in the mechanics of these two undulating swimmers. We observe that our anguilliform swimmer is able to show an over-all production of positive thrust for some parameters where the carangiform one fails to produce it within the viscous flow regime (Re = $10^2$). These parameters are shown in Table 4 for four such cases where it is evident that the pressure component of drag could not exceed the frictional part. Now, the employment of time-averaged hydrodynamic quantities alone cannot provide us with important underlying information about the behavior of undulating swimmers. An in-depth knowledge becomes more important when we conduct further comparative analysis of the hydrodynamics for our anguilliform and carangiform swimmers. For this purpose, we turn to the temporal flow fields in the close vicinity of these bodies as shown in Figure 8. We show these vorticity ($\omega$) contours for two time instants during their undulation cycles; $\frac{t}{\tau} = 0.50$ and 1.00, that reveal interesting features of the underlying governing mechanisms for hydrodynamics for the two swimmers.

These contour plots also demonstrate the strength of our computational solver in order to successfully capture the intricate details of the near-body flow fields with the swimmers undergoing large wavy oscillations. Here, we find that the timing of vortex shedding and their formation is largely affected by $\lambda^*$ as it specification dictates the temporal waveform along the body. There are smaller secondary vortices present on the bodies of both the swimmers along with larger primary vortices in different stages of their development and shedding around the trailing-edges. Around the anguilliform swimmer's body for case-1, four secondary vortices are present; two (red) on its right side and two (blue) on its left side. There exists only two secondary vortices with one present on each side of the carangiform swimmer. In case 1 of the anguilliform swimmer, we witness the birth of these secondary vortices at the anterior part of the body on its each side and they traverse along the body with its wave-speed (see **Movie 1**). The smaller blue vortex on the left side merges with the bigger blue vortex about to shed from the opposite side of the trailing-edge at $\frac{t}{\tau} = 0.50$ and the smaller red vortex on the right side meets with the bigger one coming from the left side at the trailing-edge. This constructive interference between these same-signed vortices enhances their strength while they are about to shed in the wake. The smaller vortices present near the leading-edge will undergo a similar pattern in the next undulation cycle.

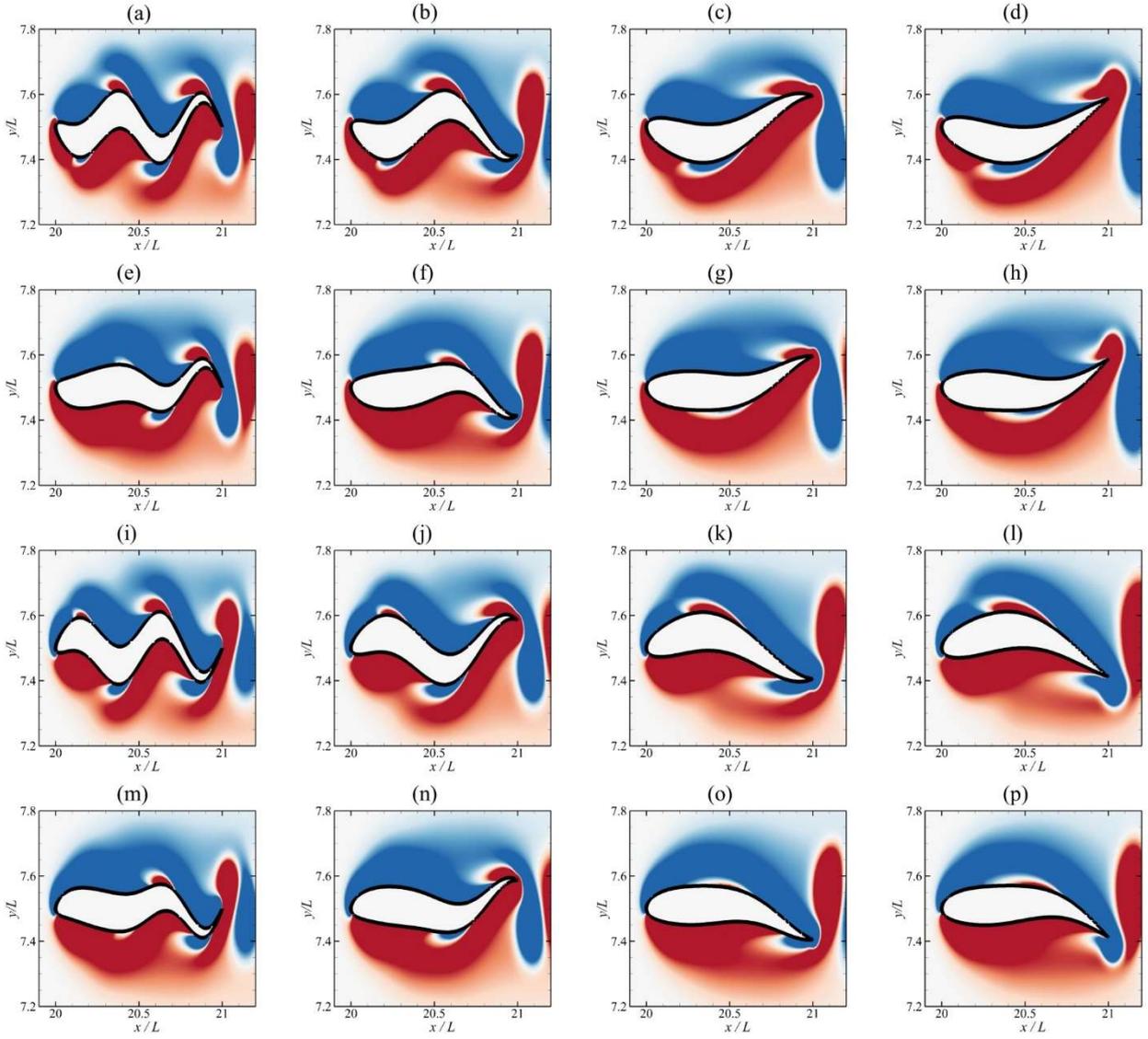

**Figure 8**: Vorticity dynamics (Vorticity contours ranging from -5; blue, to +5; red) in the close vicinity of the swimmers where the first (a – d) and second rows (e – f) shows flow fields around the anguilliform and carangiform swimmers, respectively, at $\frac{t}{\tau} = 0.50$, whereas the third (i – l) and fourth rows (m – p) shows flow fields around the anguilliform and carangiform swimmers, respectively, at $\frac{t}{\tau} = 1.00$. Here, 1st, 2nd, 3rd, and 4th columns belong to the cases 1, 2, 3, and 4, respectively as described in Table 4.

A similar dynamics is also observed around the carangiform swimmer's body, but with only two secondary vortices. These structures take birth at around one-quarter the body-length and traverse to the posterior part with the wave-speed (see **Movie 5**). During this, their sizes grow and they are shed in the wake after getting merged with the similar-signed vortices emerging from their respective opposite sides. For the other cases, both the swimmers carry two secondary vortices that have the similar fate as described previously (see **Movies 1-8**). When we compare the corresponding contours for the two bodies, their flow patterns appears to be quite resembling but with the anguilliform swimmer producing stronger secondary vortices. Hence, there may be two possible factors associated with the production of thrust within the viscous flow regime; 1) the production of more energetic primary trailing-edge vortices where the swimmer imparts more momentum to the surrounding flow due to its higher wave-speed and 2) the production of secondary vortices on the anterior parts of the bodies and their traversing to the posterior parts for their final constructive interference with their larger same-signed vortices emerging from the respective opposite sides of the trailing-edge in order to make the swimmer experience thrust or lesser drag. For lower wave-speeds, the production of secondary vortices are not evident, rather, a small portion of flow separation appears. When we increase wave-speed by increasing $f^*$, the secondary vortices start getting developed where the wave-speed helps them have a constructive interference at the trailing-edge. It is reasonable to say that the timing of this vortex interaction might also assist the BvKVS to transform into RBvKVS.

It is important to note that the maximum amplitude for the tails of both swimmers is equal that is generally considered as the characteristics length for such flow phenomena. To take it further, we can analyze the governing mechanism by computing other important flow quantities and see their trends. Figure 9 shows the temporal histories of circulation ($\Gamma$), horizontal coordinates ($X$) of the vortex centers, and phase velocities ($U_{phase}$) of positive and negative vortices in the wake of the two swimmers. We extract coordinates ($X, Y$) of the vortex centers by performing a local search in order to explore the locations of the maximum and minimum vorticity. The advection velocity of the vortices; known as the phase velocity, can be computed by numerical differentiation of their respective displacement data. Next, the circulation of a vortex is a measure of its strength and is defined as the line integral of the velocity field around its boundary or through the surface integral of vorticity field over the area of the vortex. There are different computational methods present in literature for the calculation of $\Gamma$ (Morgan, et al., 2009; Godoy-Diana, et al., 2009). To define the size of a vortex, Godoy-Diana et al. (2009) used Gaussian fits along the horizontal and vertical axes with its center on the locations of maximum and minimum vorticity and this methodology was also adopted by Zheng and Wei (2012), Zhu et al. (2014), and Wei & Zheng (2014). In this method, a vortex is encompassed by a rectangular frame with its dimensions obtained through their fitting procedure. The drawback of this method is that it may include the vorticity of the neighboring counter rotating vortices; thus increasing the chances of numerical errors in computing the circulation of a vortex. In this work, we adopt another computational methodology to find the bounding region of a vortex where we fix the bounding region of each vortex by some locations inside its neighboring counter-rotating vortices on each side. We, then, search locally for all the data points with the respective positive or negative vorticity that has the vorticity magnitude more than 5% of the maximum vorticity ($\omega_{max}$) of the flow field. The remaining data points constitute the trailing parts of vortices and their inclusion in the computations leads to numerical errors in recording the features of main vortex cores. After that, we extract the boundary points for each vortex along with the respective physical quantities of the flow field without specifying any pre-defined geometric outline. In this way, there remain no chances of the inclusion of the opposite vorticity in computing circulation. Next, we perform the numerical integration of the resultant dot product between the horizontal and vertical velocities with the relevant displacement entities. The circulation of red vortices is positive due to its counter clockwise rotation and that of the blue vortex is negative owing to its clockwise rotation.

We start recording these physical quantities for the two counter-rotating vortices by skipping the vortex attached with the trailing-edge at the beginning of an oscillation cycle for each swimmer. It is clear from Figure 9a–d that, for all the cases mentioned in Table 4, the circulation magnitude for both the positive and negative vortices produced by the anguilliform kinematics are greater than those being shed by the carangiform swimmer. It is possible to track these vortices until their vorticity magnitude remains above 0.05 times $\omega_{max}$. We also show the motion histories of these vortices in the wake in Figure 9e–h because this helps us extract $U_{phase}$; a quantity usually taken as a measure to define momentum deficit or momentum-surfeit regions in the wake. The presence of RBvKVS is associated with $U_{phase} > U_\infty$, whereas $U_{phase} < U_\infty$ is an indication of the formation of BvKVS. In Figure 9(i), we show the average $U_{phase}$ for the positive and negative vortices only at the initial time (Godoy-Diana, et al., 2009). It is evident that carangiform swimmer produces $\frac{U_{phase}}{U_\infty} < 1$ for the cases 1, 3, and 4. It produces $U_{phase} > U_\infty$ for case 2, but its consequent pressure part of the thrust force cannot overcome the frictional drag as shown in Table 4. The more important aspect is that $U_{phase}$ for the anguilliform swimmer mostly remains higher than that for the other one.

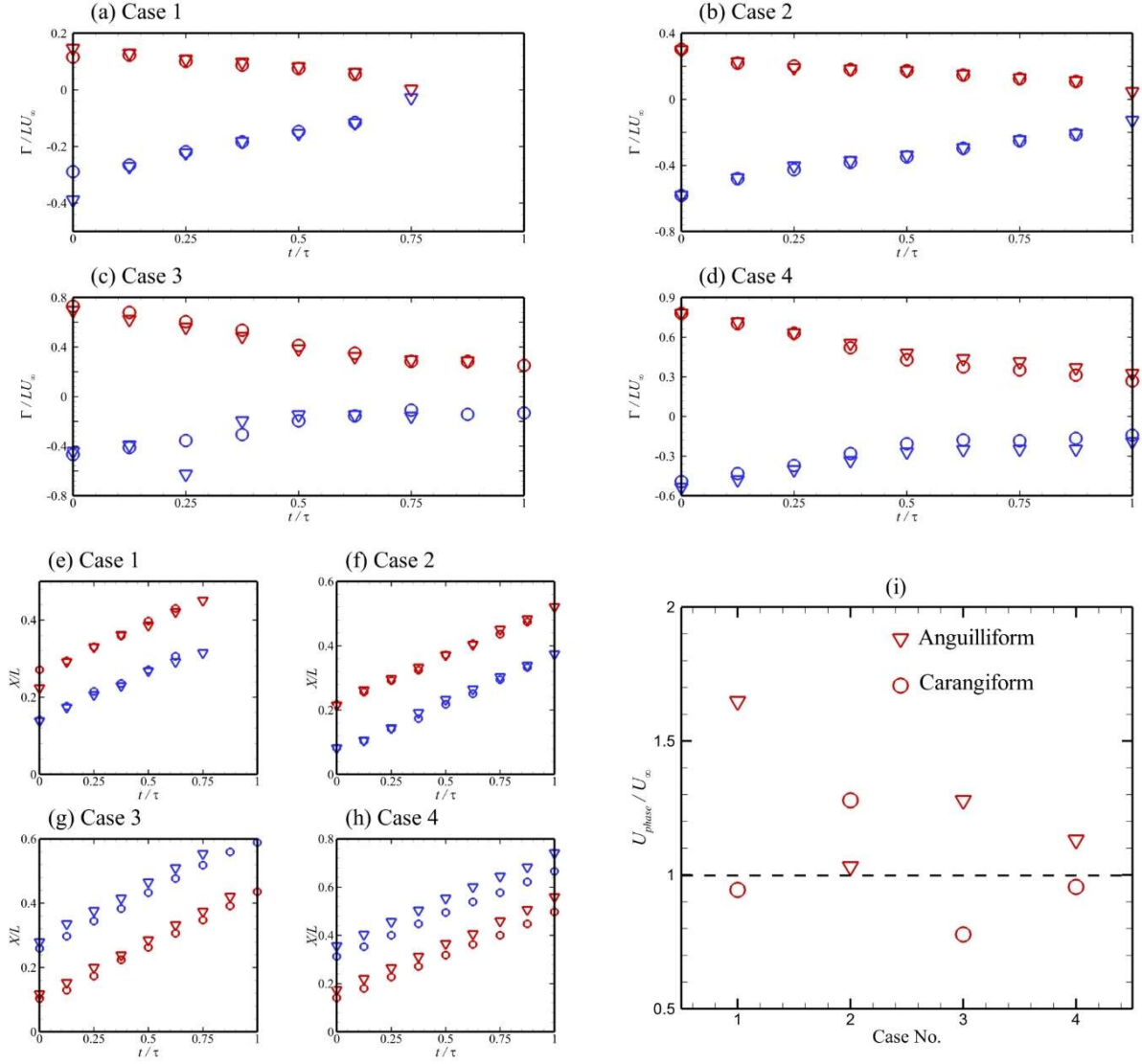

**Figure 9:** Subfigures a-d shows circulation of positive and negative vortices for our undulating swimmers. Here, red colored symbols shows data for the positive vortices, whereas blue ones are for negative vortices.

A careful inspection of these plots uncovers some other salient features about the flow physics associated with undulating swimmers. For instance, the size of the secondary vortex becomes greater with increasing $\lambda^*$. The formation and presence of these grown-up vortices explain the higher thrust values in Figure 3 for greater wavelengths. This flow characteristic may also be the reason behind the requirement of more power consumption by these swimmers to undergo their specified wavy oscillations.

Our further investigations for the near-body and wake dynamics of the swimmers for transitional and inertial flow regimes, the thickness of boundary layers around the swimmers' bodies reduces as expected. Nevertheless, near-wake dynamics shows some important distinct features for both kinds of conditions. In order to explore mechanisms governing the flow physics here, we find that almost the whole thrust production regime shows jet deflection for both anguilliform and carangiform swimmers as depicted in Figure 9. For instance, we show time-averaged flow fields in the wake of both the swimmers showing obvious jet deflections in Figure 10 at $Re = 5 \times 10^3$ and $\lambda^* = 1.00$. A very important observation here is that, contrary to those recorded for foils performing simple harmonic oscillations (Cleaver, et al., 2012; Khalid, et al., 2018; Deng, et al., 2016; Godoy-Diana, et al., 2009), and flexible filaments (Zhu, et al., 2014), the wake does not begin its deflection from the trailing-edge of the foil, but it shows its deviation at some distance from the swimmer's body while adopting a curvilinear path instead of a straight one.

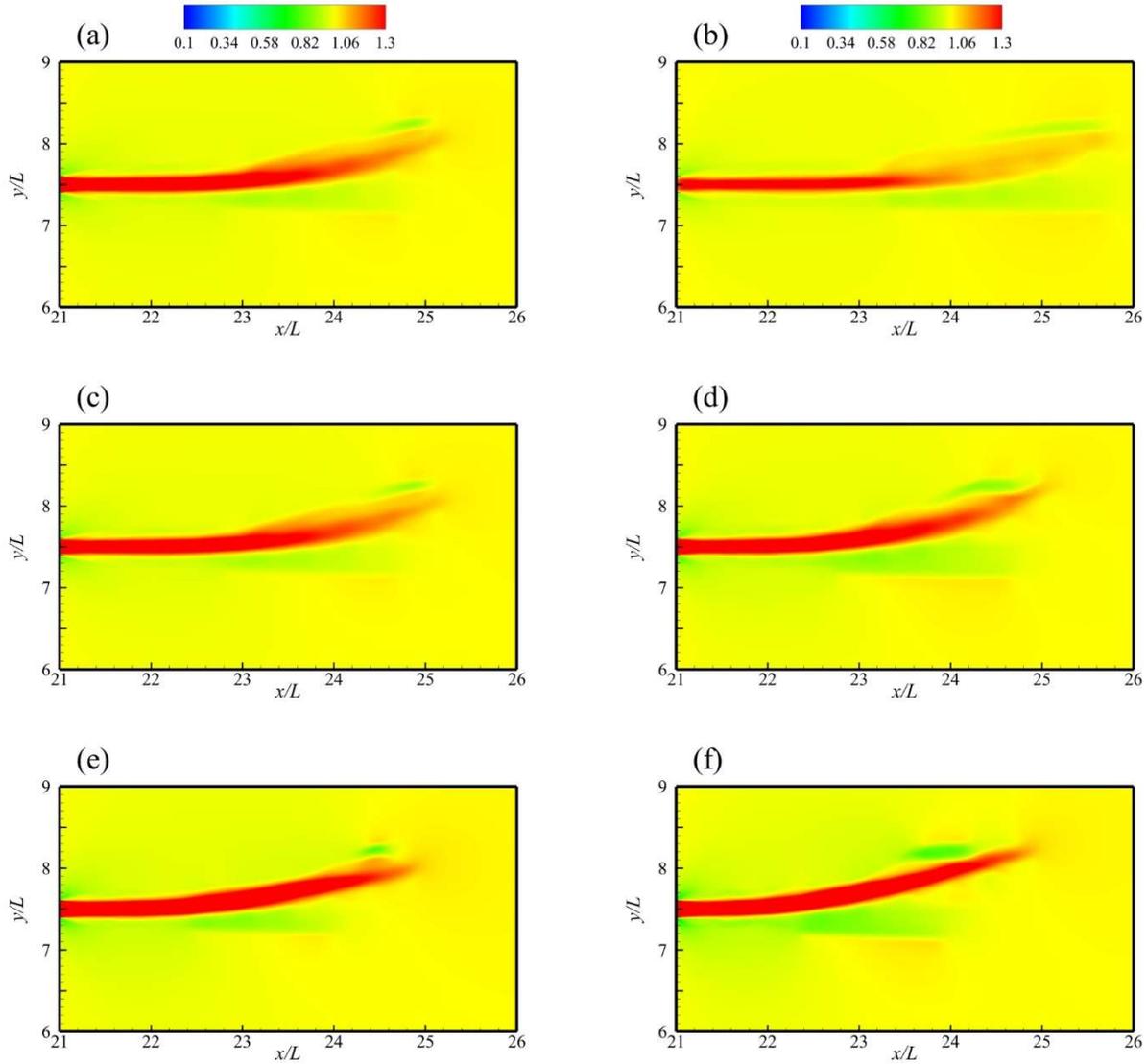

**Figure 10:** These contours show time-averaged the horizontal velocity component ($\overline{U}/U_\infty$) in the wake of the two swimmers for $Re = 5 \times 10^3$ and $\lambda^* = 1.00$. Left and right columns correspond to anguilliform and carangiform swimmers, respectively, whereas the top, middle, and bottom rows show the results for $f^* = 0.40$, 0.50, and 0.60.

As explained previously by Cleaver et al. (2012) and Khalid et al. (2015; 2018) for flows over plunging foils, such asymmetries in flows causes the emergence of non-zero time-averaged lateral force on the structure. We present time-averaged lateral force coefficients ($\overline{C_Y}$) for both the swimmers in Figure 11. It is certainly found that a greater $\lambda^*$ with a higher $f^*$ produces jet deflections and brings positive or negative non-zero $\overline{C_Y}$ depending on the direction of the deviation. A careful look at these plots reveals that there is no obvious trend in its direction of deflection. Since all the simulations start from the same temporal kinematic profile, which way jet gets deflected is not solely influenced by the initial direction of tail's stroke. Instead, this seems dependent on how vortices interact with their neighboring counterparts. For self-propelling bodies, this may result in forcing the swimmer to deviate from its straight path and experience lateral drifting unless they attempt to control the flow in their vicinity by performing some asymmetric motion. This perspective may illustrate a generic reasoning behind the asymmetric motion of dace fish (Bainbridge, 1963), eels and largemouth bass (Lauder & Madden, 2006; Xiong & Lauder, 2014), clown knifefish and bluegill sunfish (Lauder, recorded during their swimming under various flow conditions.

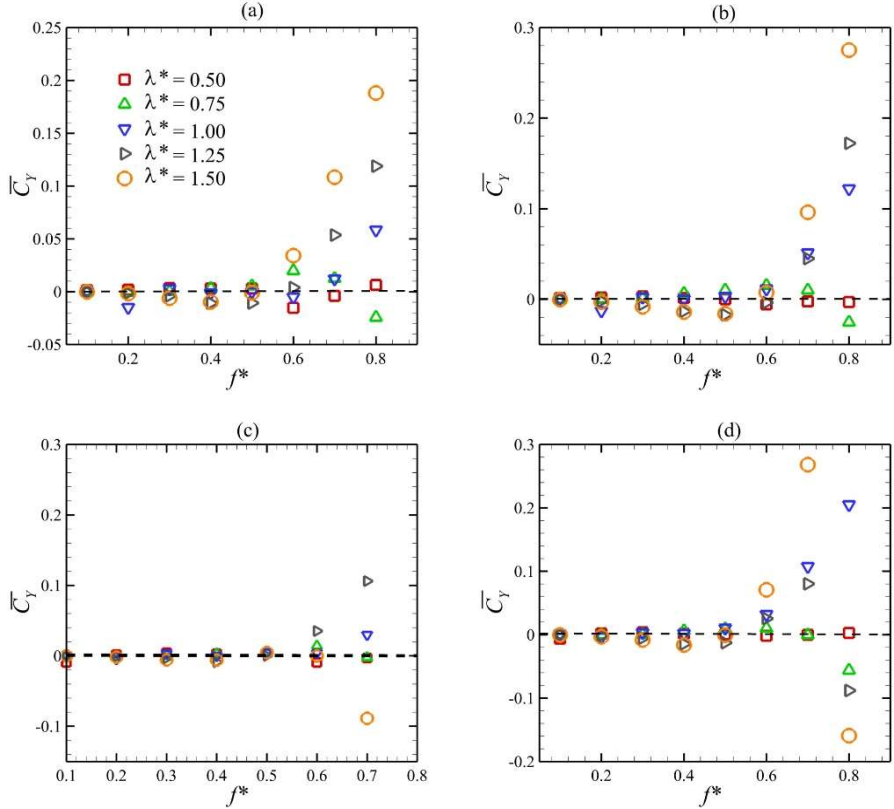

**Figure 11:** These plots present time-averaged lateral force coefficients for both the swimmers where left and right columns are for anguilliform and carangiform swimmers, respectively. The upper and lower rows show $\overline{C_Y}$ computed for flows at $Re = 10^3$ and $5 \times 10^3$.

Next, these deflecting jets in Figure 11 shows that the underlying responsible instability mechanism does not seem to initiate when the vortices are released from trailing-edges. Nonetheless, the location for the inception of the wake deflection starts getting closer to the swimmer's body for the greater $f^*$ that means a pronounced effect on the magnitude of the lateral force. Some previously presented criteria (Godoy-Diana, et al., 2009; Zheng & Wei, 2012; Wei & Zheng, 2014) for the wake deflection in case of flows over oscillating foils considered vortex dynamics to explain such phenomena. First, Godoy-Diana et al. (2009) introduced an important criterion stating that we would observe the deflection of a vortex street if the initial self-advection velocity of a dipole ($U_{dipole}$), constituted by two counter-rotating vortices shed in one oscillation cycle, exceeds the component of the phase velocity in the direction of $U_{dipole}$. It was extended to jet switching mechanisms by Zheng & Wei (2012) and Wei & Zheng (2014). This also works well for wakes behind flexible filaments at Re = 200 as reported by Zhu et al. (2014). Following these, we also hypothesize, despite some dissimilarity in the wake configurations in our case that vortex pairing mechanism becomes the responsible factor for the onset of jet deviating from the free-stream direction. Using Biot-Savart law, we introduce the translational velocity of a vortex dipole defined as follows (Safeman, 1992).

$$U_{dipole} = \Gamma/2\pi\zeta$$

where $\Gamma$ denotes the mean circulation of the counter-rotating vortices with their centers separated by the distance $\zeta$. This relation tells us that $U_{dipole}$ would increase with decreasing $\zeta$ where its little variations can bring significant changes in the estimates of $U_{dipole}$. Further guidelines taken from earlier discussions on this subject (Godoy-Diana, et al., 2009; Wei & Zheng, 2014) lead us to postulate that the jet switching process, in these cases as well, is dictated by a similar dynamics of vortex dipoles. We also hypothesize that the reason for the location for the onset of wake deviation coming closer to the body at higher $f^*$ is the provision of greater momentum to the vortices when they are released to traverse in the wake. This helps them develop stronger dipoles with their neighboring counter-rotating vortex and, consequently, dominate their phase velocity at relatively early stages.

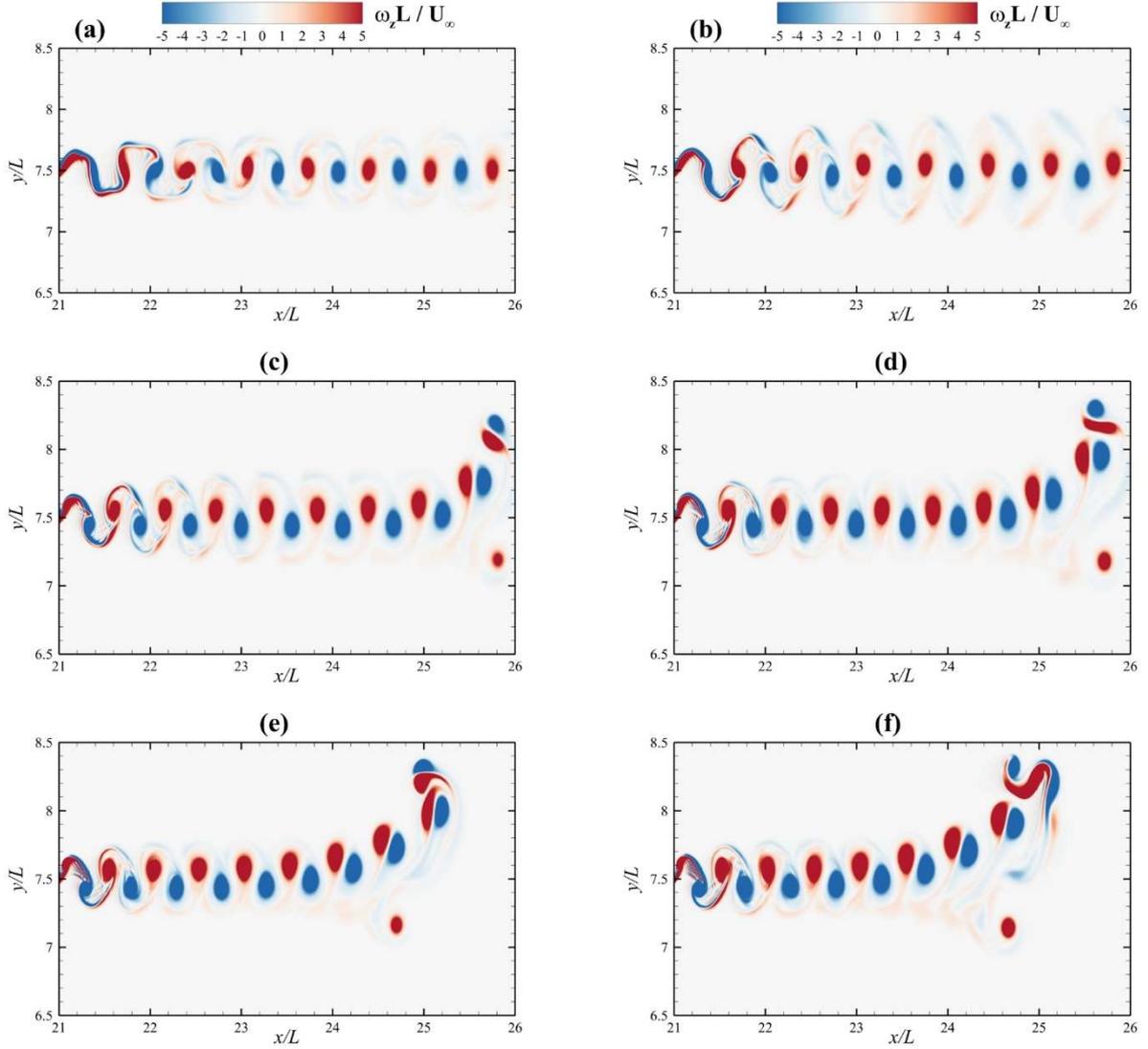

**Figure 12:** These are the vorticity contours in the wake of our undulating swimmers at the end of 10<sup>th</sup> oscillation cycle where the left and right columns show the wake dynamics of anguilliform and carangiform swimmers, respectively. These top, middle, and bottom rows are taken for $f^* = 0.30$, 0.40, and 0.50 with $\lambda^* = 1.0$ at $Re = 5 \times 10^3$.

In Figure 12, we show both kinds of cases with or without wake deflection where the vortex dynamics here clearly shows that stronger vortex interaction in the form of dipoles becomes responsible for wake deflection. With an increasing $f^*$, it appears that the required dipole formation and their velocity exceeds the phase velocity at a location closer to the bodies owing to the shedding of energetic vortices. This interaction is not seen for $f^* = 0.30$ for both the swimmers. Another significantly important aspect is the obvious connection between the optimum efficiency and wake deflection. When the onset location of jet deflection starts coming closer to the swimmer's trailing-edge, it deteriorates its $\eta$ since it forces the body to expend more amount of energy to perform undulation. In this context, when we observe the variations in $\overline{C_Y}$ in Figure 11, it is evident that, right after the kinematic parameters exceeds their optimum value to produce maximum hydrodynamic efficiency for undulating swimmers, magnitude levels of the lateral force coefficients becomes greater.

To illustrate the possible vortex pairing mechanisms to govern such phenomenon, we draw a schematic of three vortices in Figure 13; two clockwise rotating ($V_1$ and $V_3$ outlined in blue) and one counter clockwise rotating ($V_2$ outlined in red). If $V_2$ establishes its pairing with $V_1$ to form a dipole, it will have $U_{dipole}$ directed downward. Contrary to that, a pairing between $V_2$ and $V_3$ to make a dipole dipole will cause $U_{dipole}$ as shown with the upward directed arrow here. A similar analogy was also utilized by Wei and Zheng (2014) while describing the symmetry-breaking properties for a heaving airfoil. Considering the role of $\zeta$; the distance between any two neighboring counter-rotating vortices, in defining $U_{dipole}$, its smaller value will lead to stronger dipole and, consequently, a greater velocity. This Eulerian distance can be computed by $\sqrt{(X_1 - X_2)^2 + (Y_1 - Y_2)^2}$, where $X_i$ and $Y_i$ shows the rectangular coordinates for the center of a vortex. Zheng and Wei (2012) defined the following physical quantity; named as the symmetry-holding effective phase velocity in case vortex 1 and vortex 2 (see Figure 13) forms a dipole, or the symmetry-breaking effective phase velocity if vortex 2 and vortex 3 constitutes a vortex pair.

$$U_p^* = U_{dipole} - (U_{phase} - U_\infty) \cos \theta$$

Here, $\theta$ is the angle $U_{dipole}$ makes with the horizontal axis.

Unlike the previously reported phenomena (Godoy-Diana, et al., 2009; Zheng & Wei, 2012; Wei & Zheng, 2014; Zhu, et al., 2014), the wake gets deflected after covering a certain distance along the symmetry axis of the swimmer's body in case of their undulating motion. This hints towards a logical reasoning for a flow control strategy for fishes to avoid wake deflection or bring its onset location farther from their trailing parts since it degrades their swimming efficiency.

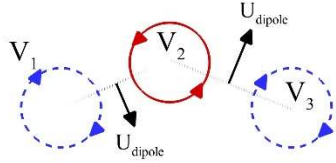

**Figure 13:** As schematic for vortex pairing mechanisms

We test these propositions by considering a few cases from our parametric space because the purpose is to check the applicability of the previously proposed criteria to the wakes of undulating swimmers which show some dissimilarity from those observed earlier. For this, we record the travel histories of the centers of the vortices and compute their circulations for a few sample cases by the methodology explained earlier in this paper. Since the averaged wake patterns seems similar for both the swimmers in Figure 10 and Figure *12*, we use our results for only the anguilliform swimmer here at Re = $5 \times 10^3$ and $\lambda^* = 1.0$ with $f^* = 0.40$, 0.50, and 0.60 to extract the characteristics of vortices in Figure 14. The choice of these data points for this particular swimmer also corresponds to the kinematics on and around the optimized efficiency production at $f^* = 0.50$ (see Figure 7).

In Figure 14a, b, and c, $X_{center}$, nondimensionalized by the body-length $L$, for all the cases seems to move in a straight line and slopes of these plots appear to be unchanged during the whole time-period. It is important to mention that we record these profiles for six oscillation cycles that provides us with significant amount of information about the wake dynamics. As $f^*$ increases, $X_{center}$ for higher Strouhal frequencies cover smaller distances from the trailing-edge of the body (zero value on the vertical axis). When we turn to the next row of these plots (Figure 14d, e, and f), there is a clear indication of wake switching its direction upwards and this initiates at earlier stages for higher $f^*$. Comparing it with the schematic in Figure 13, the trends also demonstrate that the pairing between vortex 2 (positive) and vortex 3 (negative) is stronger as compared to that between vortex 1 and vortex 2. This can be said by noticing the decreasing distance between their centers. Thus, vortex 2 and 3 form a dipole and continue their travelling in the upward direction and their pairing becomes stronger with an increasing $f^*$; a more likely situation for the jet deflection to occur. Looking at their circulation in the plots placed in the bottom row of Figure 14 (g, h, and i), their respective circulation takes higher magnitudes at the greater $f^*$ due to the shedding of more energetic vortices in the wake under these kinematic conditions. As we estimate these quantities for longer time periods, the circulations shows oscillating patterns before getting decayed due to viscous effects. It means that the wake is continuously energized by the just-shed vortices from the posterior parts of the swimmers' bodies and it counteracts the viscous effect in order to maintain the amount of vortex circulations for a longer distance. This may be the only mechanism responsible for this action against the viscous effect causing the suppressing of vortices in far-wake regions.

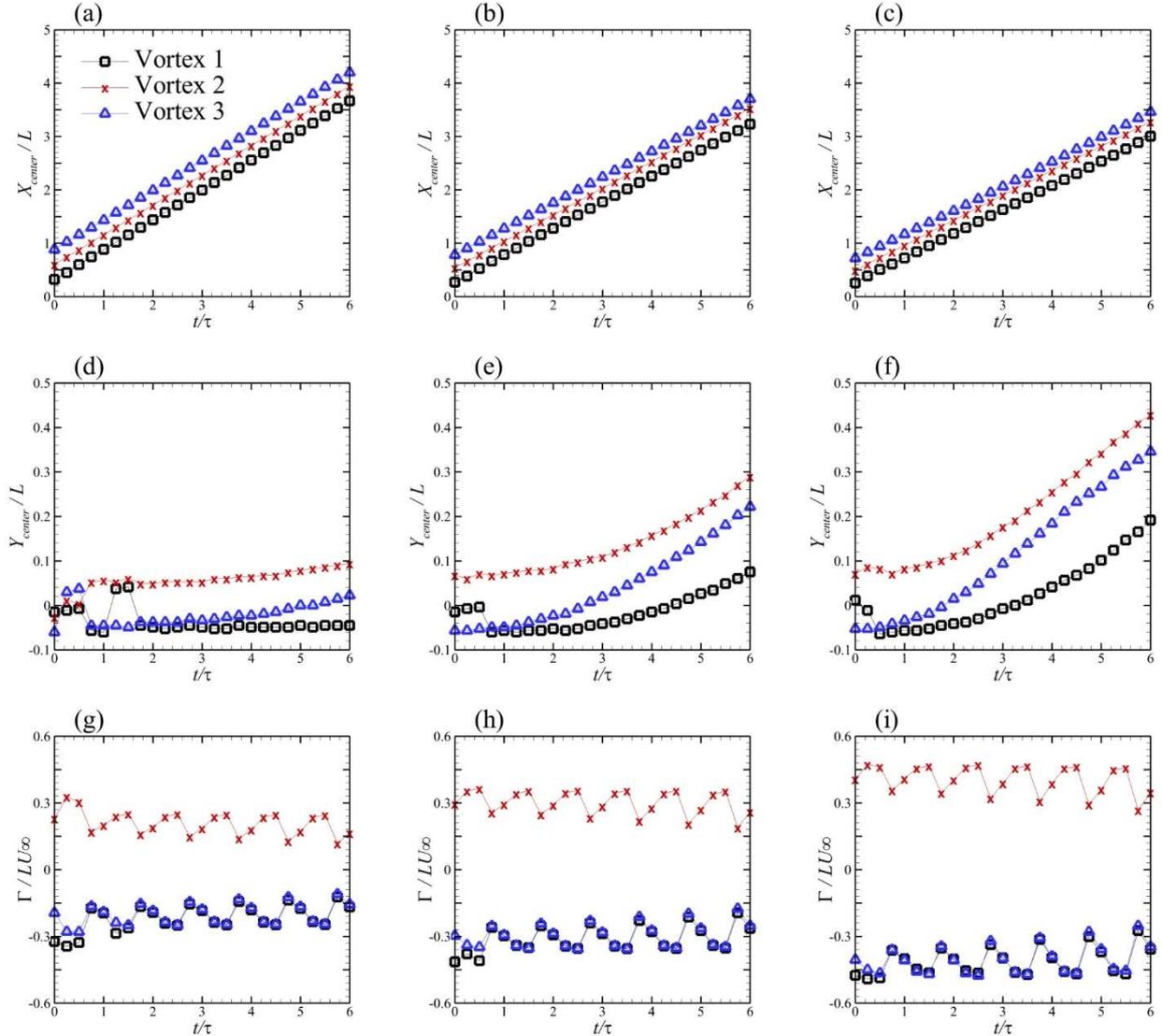

**Figure 14:** These figures provide temporal histories for coordinates of the centers of the vortices and their respective circulation in the wake of our anguilliform swimmer at $Re = 5 \times 10^3$ and $\lambda^* = 1.0$ where the left (a, d, and g), middle (b, e, and h), and right (c, f, and i) columns present data for $f^* = 0.40$, $0.50$, and $0.60$, respectively.

To illustrate the phenomenon of wake deflection, we apply the previously proposed criterion (Zheng & Wei, 2012) and compute the distances between the centers of the neighboring vortices 1, 2, and 3 along with their phase velocities responsible to direct the wake dynamics. We show this data for six oscillations cycles in Figure 15. Here, $\zeta_{12}$ and $\zeta_{23}$ denote the distance between the centers of vortices $1 - 2$ and $2 - 3$, respectively. Earlier, when such criteria were employed (Godoy-Diana, et al., 2009; Zheng & Wei, 2012; Zhu, et al., 2014), only the initial velocities were taken into account to apply this criterion. Due to the dissimilarity of the wake for undulating swimmer as explained previously, it becomes important to track the records of symmetry-breaking and holding phase velocities as the continuous energizing of wake through the swimmer's oscillations can trigger the jet switching at any time-instant while it traverses downstream. We find that the two branches in the plots of $\zeta$ starts getting separated at an earlier time. $\zeta_{23}$ decreases as we move forward in time, but $\zeta_{12}$ increases. Coming to the phase velocities, symmetry-breaking phase velocity could not exceed the symmetry-holding one for $f^* = 0.40$ and their difference remains minimal, thus not allowing the switching of jet in the upward direction.

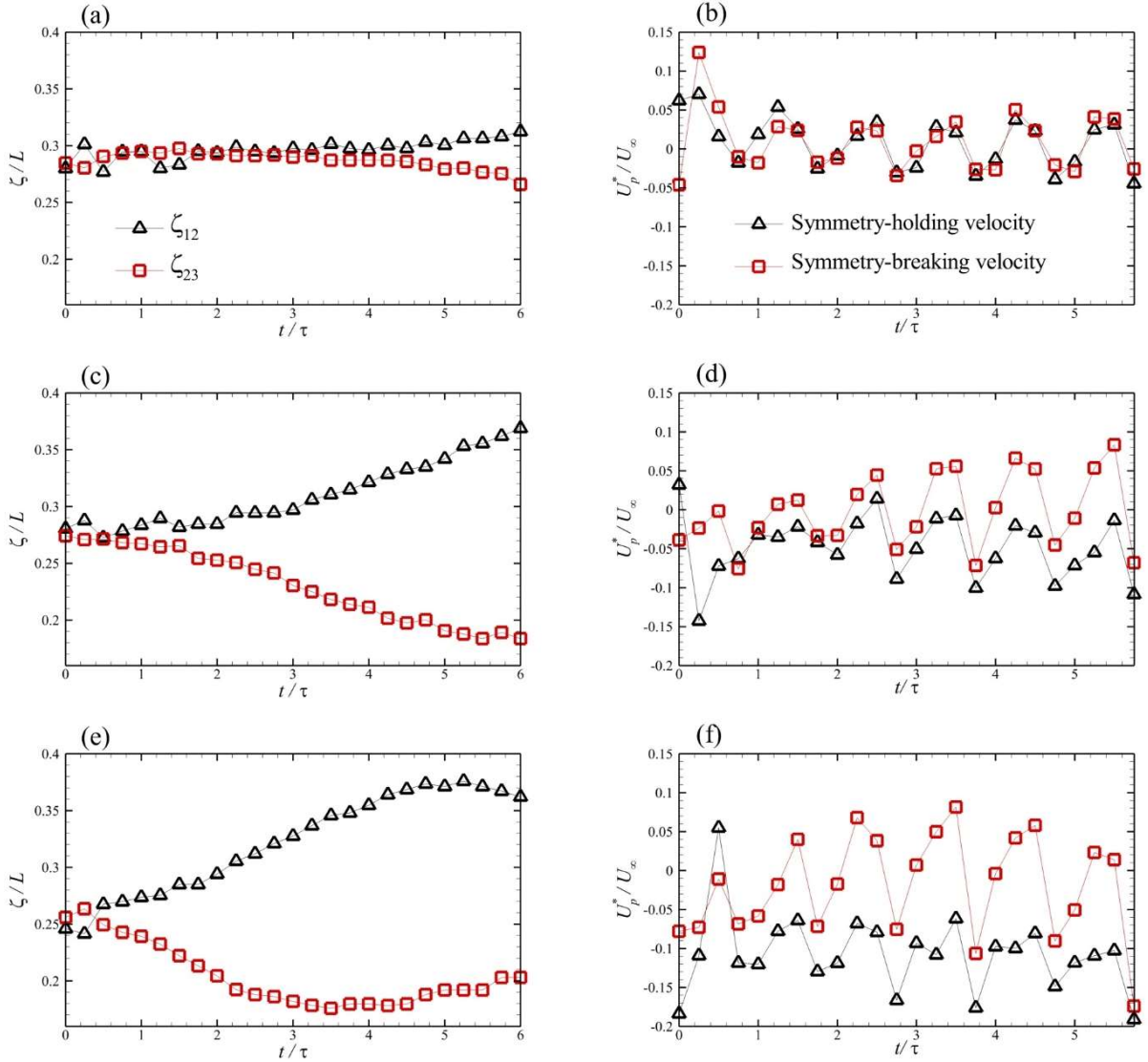

**Figure 15:** These plots (a, c, and e) show the distance between the centers of vortices forming dipoles 1 and 2, whereas the plots in the right column present symmetry-breaking and symmetry-holding velocities. Top, middle, and bottom rows are for $f^* = 0.40$, 0.50, and 0.60.

For $f^* = 0.50$, the symmetry-breaking velocity exceeds the other one at $\frac{t}{\tau} > 1$, but it happens at $\frac{t}{\tau} < 1.0$ for $f^* = 0.60$; an indicator for the onset of wake deflection closer to the trailing-edge of the swimmer. Hence, these measurements support our propositions stated in the beginning of our discussion on this subject.

## 4. SUMMARY & CONCLUSIONS

In this paper, we perform numerical simulations for a wide range of flow and kinematic conditions considering Reynolds number, Strouhal frequency, and wavelength as the governing parameters to investigate important issues pertaining to the hydrodynamics of anguilliform and carangiform swimmers. We find that the anguilliform swimmer is a better candidate for viscous flow regimes Using physical characteristics, such as the center and circulation, of vortices, we explain how an anguilliform swimmer can do this. For both the swimmers, we find the constructive interference of secondary vortices traversing from anterior parts of bodies with ready-to-shed primary trailing-edge vortices formed at the other side of the swimmer's body. It appears that this secondary interference becomes responsible for producing greater circulation of vortices and help them form reverse Benard von Karman vortex street. Also, our anguilliform swimmer's hydrodynamic performance at higher Reynolds numbers dominates that of carangiform swimmer in terms of both thrust production and efficiency for $\lambda^* \geq 1.0$ for the whole range of $f^*$. However, the carangiform kinematics is an obvious choice if it needs to swim with $\lambda^* < 1.0$. We also compare and report their individual performances for a range of $f^*$ and $\lambda^*$.

We show that most of the thrust production regimes in transitional and inertial flows is suffered by the wake deflection. Our results show that the hydrodynamic efficiencies of both the anguilliform and carangiform swimmers reach to their respective maximum values just before the wake deflection begins impacting the lateral force magnitudes and, from that point onward, the swimmers need to consume more energy to perform their wavy motion. We believe that, in order to circumvent this problem and keep the vortices aligned in the thrust force direction, the fish attempts to perform asymmetric undulatory motion that was recorded by many researchers previously. Since, we are currently investigating this phenomenon to settle this issue and the subject is not inside the scope of our present paper, it will be presented to the scientific community in the near future.

We have also observed that the wavelengths usually adopted by natural swimmers do not lead them to get most of the hydrodynamic advantage. Contrary to that, anguilliform kinematics of a body can outperform the carangiform swimmers if it swims with greater wavelength of the wave along its body. Similarly, the carangiform motion can hydrodynamically dominate its counterpart if it utilizes lower wavelengths. Quite astonishingly, we find an entirely opposite pattern in the motion of natural aquatic species. Based on our detailed quantification, It may be concluded that the fish physiology, for both anguilliform and carangiform swimmers, does not support them to attain the maximized hydrodynamic advantages and they attempt to optimize their propulsive characteristics based on their respective morphologies.

As we clearly find a qualitative connection between the wake deflection and the optimum kinematic parameters of undulatory swimmers, we observe that the wakes behind these swimmers deflects while taking a curvilinear path; unlike those of the foils undergoing simple harmonic oscillations that gets deflected right at the releasing time of the vortices at their trailing edges. Nevertheless, the condition of the symmetry-breaking velocity exceeding the symmetry-holding velocity still holds. The major difference lies in the fact that it can happen when the vortices traverse some distance from the bodies and the instability can be triggered by the energy being fed in by the shedding of vortices in the wake. We support this argument by showing some sample temporal profiles of circulation for vortices. The oscillating time-histories show that the viscous effect in the wake of these undulating swimmers at intermediate and inertial flow conditions is opposed by the swimmers by injecting more energy in the wake and providing a push to the vortices already traversing at some distance. This logical reasoning implies that the amount of this energy provided by the swimmer needs to be controlled in such a way that the symmetry-breaking phase velocity of a dipole does not exceed the symmetry-holding phase velocity somewhere closer to the body. If it happens farther from the body, it is less likely to impact the swimming efficiency of a fish. However, if the fish is compelled to swim in such circumstances, it may adopt some other mechanisms, such as asymmetric undulation, to keep the trail of vortices on track as much as possible. This statement comes from our conventional wisdom and needs further scientific investigations that we aim to present soon.

**Declaration of Interests**

The authors do not have any known competing financial interests or personal relationships that could appear to have influenced our work reported in this paper.

**Acknowledgements**

First author is a recipient of International Exchange Research Fellowship sponsored by Chinese National Science Foundation and Peking University, Beijing.